\documentclass{PoS}

\title{Topology of two-color QCD at low temperature and high density}

\ShortTitle{Topology of two-color QCD at low temperature and high density}

\author{\speaker{Etsuko Itou}\\
        Department of Physics, and Research and Education Center for Natural Sciences, Keio University, 4-1-1 Hiyoshi, Yokohama, Kanagawa 223-8521, Japan\\
        Department of Mathematics and Physics, 
        Kochi University, Kochi 780-8520, Japan\\
        Research Center for Nuclear Physics (RCNP), Osaka University, Osaka 567-0047, Japan\\
        E-mail: \email{itou@yukawa.kyoto-u.ac.jp}}

\author{{Kei Iida}\\
 Department of Mathematics and Physics, 
Kochi University, Kochi 780-8520, Japan\\
E-mail: \email{keiiida@riken.jp}}
        
 \author{{Tong-Gyu Lee}\\
 Department of Mathematics and Physics, 
Kochi University, Kochi 780-8520, Japan\\
 E-mail: \email{tonggyu.lee@yukawa.kyoto-u.ac.jp}}

\abstract{
The chemical potential ($\mu$) dependence of the topological susceptibility with two-color two-flavor QCD is studied.  
We find that at temperature $T \approx T_c /2$, where $T_c$ denotes the critical temperature at zero chemical potential, the topological susceptibility is almost constant throughout $0\leq  a\mu \lesssim 1.0$, while at $T \approx T_c$, it decreases significantly from the $\mu=0$ value in a high $\mu$ regime.  
 
In this work, we perform the simulation for $\mu/T \le 16$, which covers even the low temperature and the high chemical potential regime.  In this regime, we introduce a diquark source term, which is characterized by $j$, into the action. 
We also show our results for the phase diagram in a low temperature 
regime ($T \approx T_c/2$), which is obtained after taking the $j \to 0$ limit of the diquark condensate and the Polyakov loop.
}

\FullConference{The 36th Annual International Symposium on Lattice Field Theory - LATTICE2018\\
		22-28 July, 2018\\
		Michigan State University, East Lansing, Michigan, USA.}

\usepackage{amsmath}
\newcommand{\beq}{\begin{eqnarray}}
\newcommand{\eeq}{\end{eqnarray}}

\begin{document}

\section{Introduction: two-color QCD with diquark source}
To elucidate the QCD phase diagram in the low-temperature and high-density regime is still a hard task, since there are at least two principle difficulties, namely, the sign problem and the numerical instability problem.
In this work, we avoid the sign problem by considering the SU(2) $N_f=2$ theory.
The fundamental representation (=quark) of the SU($2$) gauge group takes a (pseudo)real representation,
so that the SU(2) gauge theory coupled to an even number of flavors does not suffer from the sign problem~\cite{Muroya:2000qp}.
On the other hand, numerical instability occurs in the high-density regime ($\mu \gtrsim  m_{PS}/2$ with the quark chemical potential $\mu$ and the pseudoscalar meson mass $m_{PS}$).
This instability comes from a dynamical pair-creation and/or pair-annihilation of the lightest hadrons.  To solve this problem we have to modify the action~\cite{Kogut:2002cm}.
In this work, we introduce a diquark source term in the action by utilizing the Wilson lattice fermion~\cite{Hands:2006ve} and then study the phase structure and its topological property in the low-temperature and high-density regime that satisfies
$\mu/T \le 16$ with the temperature $T$.

\section{Simulation detail}
The lattice gauge and fermion actions used in this work are the Iwasaki gauge and the naive Wilson actions, respectively. 
The two-flavor fermion action including the quark number operator and the diquark source term on the lattice is given by
\beq
S_F= \bar{\psi}_1 \Delta(\mu)\psi_1 + \bar{\psi}_2 \Delta(\mu) \psi_2 - J \bar{\psi}_1 (C \gamma_5) \tau_2 \bar{\psi}_2^{T} + \bar{J} \psi_2^T (C \gamma_5) \tau_2 \psi_1.\label{eq:action}
\eeq
Here, the indices $1,2$ denote the flavor label, and $J$ and $\bar{J}$ correspond to the anti-diquark and diquark source parameters, respectively.
For simplicity, we put $J=\bar{J}$ and assume that it takes a real value. Note that $J=j \kappa$, where $j$ is a source parameter in the continuum theory, and the factor $\kappa$ comes from the redefinition of the Wilson fermion on the lattice.
The operator $C$ in the last two terms denotes the charge-conjugation operator, and $\tau_2$ acts on the color index.

The action (\ref{eq:action}) includes three types of fermion bilinears: $\bar{\psi} \psi$,$\bar{\psi} \bar{\psi}$, and $\psi \psi$. 
To write the kernel matrix in the fermion action in a single form, we introduce 
the extended fermion matrix ($\mathcal M$) as
\beq
S_F&=& (\bar{\psi} ~~ \bar{\varphi}) \left( 
\begin{array}{cc}
\Delta(\mu) & J \gamma_5 \\
-\bar{J} \gamma_5 & \Delta(-\mu) 
\end{array}
\right)
\left( 
\begin{array}{c}
\psi  \\
\varphi  
\end{array}
\right)
 \equiv  \bar{\Psi} {\mathcal M} \Psi,  \label{eq:def-M}
\eeq
where
$\bar{\varphi}=-\psi_2^T C \tau_2,$ and $\varphi=C^{-1} \tau_2 \bar{\psi}_2^T.$
The square of the extended matrix can be diagonal if $J=\bar{J}$ takes a real value. Then, we obtain
\beq
\det[{\mathcal M}^\dag {\mathcal M}] = 
\det[ \Delta^\dag(\mu)\Delta(\mu) + J^2 ] \det[  \Delta^\dag (-\mu) \Delta(-\mu) + J^2  ]. \label{eq:MdagM}
\eeq
Note that $\det[{\mathcal M}^\dag {\mathcal M}]$ corresponds to the fermion action for the {\it four-flavor} theory, since a single $\mathcal{M}$ in Eq.(\ref{eq:def-M}) expresses the fermion kernel of the two-flavor theory.
To reduce the number of fermions, we take the root of the extended matrix in the action. In dealing with this root, we utilize the Rational Hybrid Montecarlo (RHMC) algorithm in our numerical simulations.

\section{Results}
Before investigating the finite density regime, we have fixed our lattice parameters at $\mu=0$. The lattice extent in the $\mu=0$ simulations is set to $16^3 \times 32$ and $24^3 \times 48$~\cite{Lee}. The hopping parameter ($\kappa$) for each $\beta$ is tuned in such as way as to keep $m_{PS}/m_{V}=0.8$, where $m_V$ is the vector meson mass. 
In fact, the calculations with $(\beta, \kappa)=(0.8, 0.1590), (1.0, 0.1470)$ give $(m_{PS}/m_V, am_{PS})=(0.809(14), 0.619(4)), (0.795(13) , 0.368(7))$, respectively.
Utilizing the reference scale ($t_0$ scale) in the gradient-flow method, we have also obtained the relationship between $\beta$ and the lattice spacing.
The lattice spacing at $\beta=1.0$ is almost half of the one at $\beta=0.8$.

The Hadronic-QGP phase transition at $\mu=0$ occurs around $\beta=0.8$ and $N_\tau=8$.   In this work, we thus denote the temperature at $\beta=0.8$ and $N_\tau=8$ as $T_c$.  We show the results obtained for the fixed 
$\beta$ ($0.8$ and $1.0$) on the lattice extent ($16^4$).
In the common notation, they correspond to the ``zero-temperature" simulations, but here we manifestly express the corresponding temperatures as $T=T_c/2$ and $T=T_c$, respectively.

\subsection{Phase diagram at $T = T_c/2$}
Firstly, we determine the finite $\mu$ phase diagram at $T=T_c/2$.
We focus on two order parameters,  namely, the Polyakov loop and the diquark condensate.
The magnitude of the Polyakov loop ($|L|$) is an approximate order parameter for confinement, while the diquark condensate ($\langle qq \rangle \equiv \frac{\kappa}{2} \langle \bar{\psi}_1 C\gamma_5 \tau_2 \bar{\psi}_2^T - \psi_1 C \gamma_5 \tau_2 \psi_2^T \rangle
$) is the one for superfluidity.
Following the previous works~\cite{Kogut:2002cm,Hands:2006ve,Braguta:2016cpw}, we use the name of each phase as shown in Table~\ref{table:phase}.
\begin{table}[h]
\begin{center}
\begin{tabular}{|c||c|c|c|c|}
\hline
 &    Hadronic  & Quark-Gluon-Plasma & Bose-Einstein-Condensate & Superfluid \\ 
 &    & (QGP) & (BEC) & (SF) \\
 \hline \hline
$\langle |L| \rangle$ & zero  & non-zero & zero & non-zero \\
$\langle qq \rangle$ & zero  & zero & non-zero & non-zero \\ 
 \hline
\end{tabular}
\caption{ Definition of the phases. } \label{table:phase}
\end{center}
\end{table}

\begin{figure}[h]
\centering\includegraphics[width=11cm]{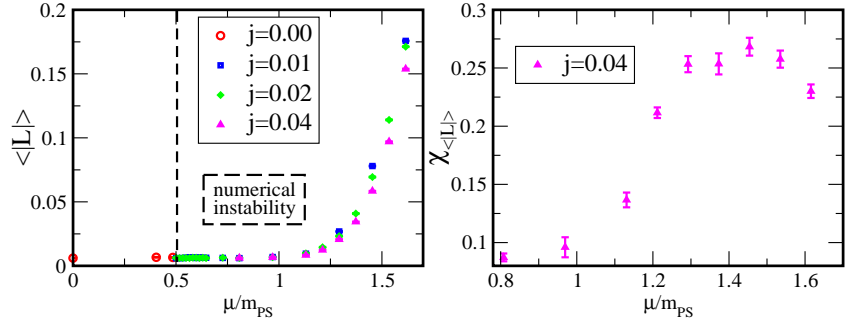}
\caption{(Left) The $\mu$ and $j$ dependences of the Polyakov loop at $\beta=0.8$ on the $16^4$ lattices. (Right) The susceptibility of the Polyakov loop with the $j=0.04$ simulations.  }
\label{fig:Ploop}
\end{figure}

The $\mu$-dependence of the Polyakov loop is shown in Fig.~\ref{fig:Ploop}.
The values of the diquark source ($j=J/\kappa$) are distinguishable by color; 
the data with red, blue, green, and magenta symbols are generated by the $j=0.00, 0.01, 0.02$, and $j=0.04$ simulations.
Note that the theoretical threshold $\mu$ value of the numerical instability, where the lightest hadrons are frequently created and annihilated in the medium, is $\mu/m_{PS}=1/2$.
In fact, we can carry out the HMC simulation without the diquark source for $\mu/m_{PS} < 1/2$, but the Metropolis test in the HMC simulation can not be accepted even within a tiny Montecarlo step ($\sim 1/1000$) for $\mu/m_{PS} \gtrsim 1/2$.
In such a high $\mu$ regime, therefore, we utilize the RHMC algorithm including the diquark source in the action.

The left panel of Fig.~\ref{fig:Ploop} shows that the value of the Polyakov loop is non-zero for $\mu/m_{PS} \gtrsim 1.2$.
It is a signal of the confinement/deconfinement phase transition (or crossover) in the high density regime.
The susceptibility of the Polyakov loop obtained at $j=0.04$ is exhibited in the right panel of Fig.~\ref{fig:Ploop}.
There is a peak at $\mu_D /m_{PS}\sim 1.45$.
No clear $j$-dependence of the peak position is observed in our calculations.

The second quantity to be calculated in determining the phase diagram is the expectation value of the diquark condensate.
The difficulty in finding the critical $\mu$ where superfluidity emerges comes from the extrapolation of the $j\rightarrow 0$ limit.
We propose a reweighting method with respect to $j$ at fixed $\beta,\mu$, and $\kappa$ ~\footnote{A similar reweighting method is discussed in the context of QCD with the isospin chemical potential~\cite{Brandt:2017oyy}.}.
Then, the reweighting factor, generally given between the original lattice parameter set ($\beta_0,\kappa_0,\mu_0,j_0$) and the measured parameter set ($\beta,\kappa,\mu,j$), reads
\beq
R_j &\equiv&  \left[ \frac{\det [D (\kappa,\mu,j)]}{\det [D (\kappa_0,\mu_0,j_0)]} \right] ^{1/2} e^{-(\beta-\beta_0) S_g[U]} \nonumber\\
&=&\det [1+ (J^2-J_0^2)(\Delta^\dag (\mu) \Delta(\mu)+J_0^2)^{-1})]^{1/2}  \det [1+ (J^2-J_0^2) (\Delta^\dag (-\mu) \Delta(-\mu)+J_0^2)^{-1})]^{1/2}. \nonumber
\eeq
In our calculation, the value of $J=j\cdot \kappa$ is at most $O(10^{-6})$, so that the Taylor expansion, $\det [1+A] = e^{ \mbox{ Tr} \ln [1+A] } \sim 1+\mbox{Tr} [A]$, works very well.
In fact, we numerically confirm that the leading correction of the reweighting factor ($R_j-1$) is less than $O(10^{-5})$ in our calculations.

\begin{figure}[h]
\vspace{0.1cm}
\centering\includegraphics[width=11cm]{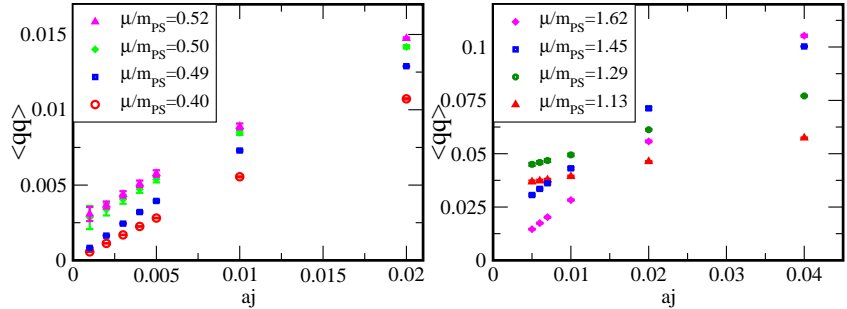}
\caption{The $j$-dependence of the diquark condensate for several $\mu/m_{PS}$.}
\label{fig:diquark}
\end{figure}
The left and right panels in Fig.~\ref{fig:diquark} present the $j$-dependence of the diquark condensate in the low-$\mu$ and high-$\mu$ regimes, respectively.
The data for $j < 0.01$ are obtained by the reweighting from the configurations generated at $j=0.01$.

The left panel shows that the data denoted by red-circle and blue-square symbols ($\mu/m_{PS} < 1/2$) go to zero in the $j\to 0$ limit, while those by green-diamond and magenta-triangle symbols ($\mu/m_{PS} \ge 1/2$) have non-zero expectation values in the same limit.
The critical value, $\mu_B/m_{PS}=1/2$, is consistent with the threshold value of the numerical instability in the HMC simulation without the diquark source in the action.
It indicates the appearance of superfluidity in the calculation without the $j$-term. 

In the high $\mu$ regime, as shown in the right panel in Fig.~\ref{fig:diquark}, we find that the $\mu$-dependence of the diquark condensate changes from the $j=0.04$ behavior in the smaller $j$ regime. 
In fact, the extrapolated value in the $j\to 0$ limit decreases with $\mu$ for $\mu/m_{PS} \gtrsim 1.45$, which corresponds to $a\mu \gtrsim 0.90$ in lattice unit.
The same behavior has been observed in the SU($2$) $N_f=4$ staggered case in Ref.~\cite{Kogut:2002cm} and has been suggested to be a signal of the lattice artifact.
We need a careful study of the volume- and mass-dependence to give a conclusion.

\begin{figure}[h]
\centering\includegraphics[width=11cm]{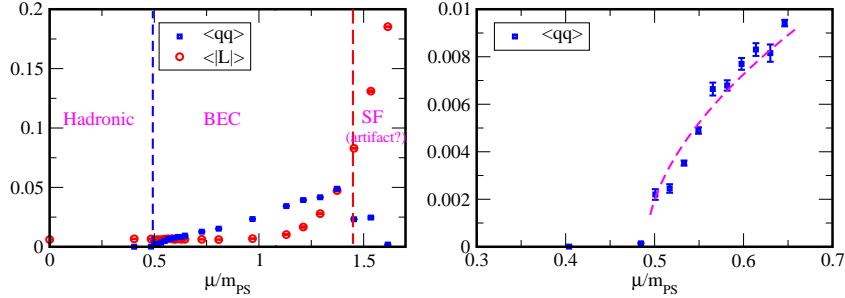}
\caption{(Left) Summary of the data of the Polyakov loop and the diquark condensate in the $j\to 0$ limit. (Right) The data of the diquark condensate in the $j\to 0$ limit around the critical point ($\mu_B$) and its fitting function.}
\label{fig:phase-diagram}
\end{figure}

Now, let us summarize the data of the Polyakov loop and the diquark condensate,
which give a phase diagram shown in Fig.~\ref{fig:phase-diagram}.
According to our notation of each phase given in Table~\ref{table:phase}, there are three phases at $T=T_c/2$, namely, the hadronic, BEC, and SF phases.
Thanks to reweighting in $j$, we can precisely find the location of the phase transition between the hadronic and BEC phases as $\mu_B/m_{PS} \approx 1/2 $. 
The scaling law around the critical point,
\beq
\langle qq \rangle \propto (\mu -\mu_B)^{\beta_m},
\eeq
is also investigated.
Here, we take $\mu_B/m_{PS}=0.49$ and obtain the best-fit value of the critical exponent as $\beta_m =0.546$; the corresponding scaling is shown by dashed curve (magenta) in the right panel in Fig.~\ref{fig:phase-diagram}. 
It is roughly consistent with the prediction by the chiral perturbation theory in the mean-field approximation ($\beta_m=1/2$)~\cite{Kogut:2000ek}.

\subsection{Topological charge at finite density}
Now, let us study the $\mu$-dependence of the topological charge (instanton number).
For this purpose, we utilize the gluonic definition:
\beq
Q = \frac{1}{32 \pi^2} \sum_{x} \epsilon_{\mu \nu \rho \sigma} F_{\mu \nu} F_{\rho \sigma} (x),
\eeq
and use the gradient flow method in measuring $Q$.
The number of measured configurations for each set of the lattice parameters is $50$, and the flow time of the gradient flow is also fixed at $t/a^2=40$.

\begin{figure}[h]
\centering\includegraphics[width=11cm]{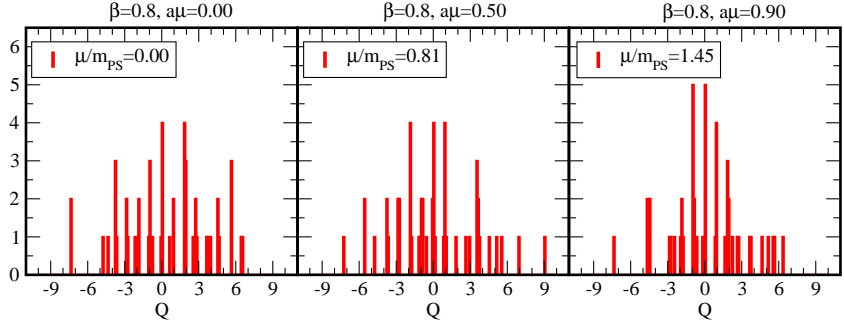}
\caption{Histogram of the topological charge in the lattice with $\beta=0.8$ and $N_s=N_\tau=16$ ($T\approx T_c/2$).}
\label{fig:hist-Q-beta0.8}
\end{figure}

Figure~\ref{fig:hist-Q-beta0.8} presents the histogram of the topological charge.
The left, middle, and right panels show the data obtained at $\mu/m_{PS}=0.00, 0.81$, and $1.45$ ($a \mu=0.00,0.50,$ and $0.90$), respectively. 
Here, the data in $\mu > 0$ are obtained by the RHMC simulations with $j=0.04$. 
They are typical of the hadronic, BEC, and SF phases.
No clear difference can be found among them.
In fact, the topological susceptibility in each phase is summarized in Table~\ref{table:topo}.
\begin{table}[h]
\begin{center}
\begin{tabular}{|c||c|c|c|}
\hline
      &    $a\mu=0.00$  & $a\mu=0.50$ & $a\mu=0.90$\\
 \hline \hline
$\beta=0.8$ & $2.85(52)\times 10^{-3}$ & $2.65(48)\times 10^{-3}$ & $2.08(31)\times 10^{-3}$ \\
$\beta=1.0$ & $4.80(127)\times 10^{-4}$ & & $4.21(132)\times 10^{-5}$\\
 \hline
\end{tabular}
\caption{ Topological susceptibility ($\chi_Q$) for various sets of the lattice prameters. } \label{table:topo}
\end{center}
\end{table}
They agree with each other within 2 $\sigma$.
We also confirm that the results obtained with $j=0.02$ for each $\mu (>0)$ are consistent with the corresponding data with $j=0.04$.

The result is qualitatively different from the earlier works that used the SU(2) $N_f=4$ theory on $12^3 \times 24$ lattice~\cite{Hands:2011hd} and SU($2$) $N_f=8$ theory on $14^3 \times 6$ lattice~\cite{Alles:2006ea}.
In these works, the decrease in the topological susceptibility with $\mu$ has been observed in the high $\mu$ regime.
To find a hint on the origin of such qualitative difference, we also investigate the $\beta$ (or $T$) dependence of the topological susceptibility.

\begin{figure}[h]
\centering\includegraphics[width=11cm]{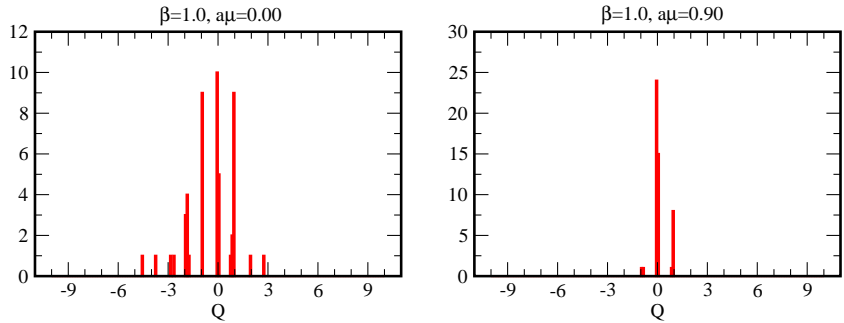}
\caption{Histogram of the topological charge in the lattice with $\beta=1.0$ and $N_s=N_\tau=16$ ($T\approx T_c$).}
\label{fig:hist-Q-others}
\end{figure}
Figure~\ref{fig:hist-Q-others} shows the histogram of $Q$ obtained at $\beta=1.0$, but with the same mass-ratio ($m_{PS}/m_V=0.8)$ and lattice extent ($16^4$).
At $\mu=0.00$, there is still a broad distribution of $Q$, while in the high chemical potential regime ($a\mu=0.90$), most configurations localize at the $Q=0$ sector.
It suggests that the $\mu$-dependence of the topological susceptibility strongly depends on the temperature; near $T=T_c$ it decreases from the $\mu=0$ value in the high $\mu$ regime, but at the lower temperature ($T \approx T_c/2$) it becomes independent of $\mu$.

\section{Summary}
We investigate the phase diagram at low temperature ($T\approx T_c/2$) and find that there are at least three phases: the hadronic, BEC, and superfluid phases.
To determine the critical point at which superfluidity emerges, we propose the method for reweighting in the diquark source parameter while keeping the other lattice-parameters fixed.
The critical exponent of the diquark condensate is almost consistent with the prediction from the chiral perturbation theory in the leading approximation.
Furthermore, we find that the diquark condensate in the $j\to 0$ limit decreases with $\mu$ near $a \mu =1.0$. It might be manifestation of a lattice-artifact phase by the saturation of the baryon density on the finite lattice.

We also investigate the topological susceptibility in all three phases.
At $T\approx T_c/2$, the topological susceptibility is independent of the value of the chemical potential for $a\mu < 1.0$.
Near $T=T_c$, on the other hand, it drastically decreases from the $\mu=0$ value in the high chemical potential regime.
We expect that a transition from the superfluid to the QGP phase takes place in the high $\mu$ regime between $T \approx T_c/2$ and $T\approx T_c$.

\section*{Acknowledgment}
\noindent We are grateful to S.~Aoki, K.~Fukushima, P.~de~Forcrand, K.~Nagata, A.~Ohnishi, and N.~Yamamoto for useful comments.
The numerical simulations were carried out on SX-ACE and OCTOPUS
 at Cybermedia Center (CMC) and Research Center for Nuclear Physics (RCNP), Osaka University,
 together with XC40 at Yukawa Institute for Theoretical Physics (YITP)
 and Institute for Information Management and Communication (IIMC), Kyoto University.
This work has partially used computational resources of
 HPCI-JHPCN System Research Project (Project ID: jh180042) in Japan.
This work is supported by Sasakawa Grants for Science Fellows (SGSF), THE JAPAN SCIENCE SOCIETY.
This work is also supported in part by the Ministry of Education, Culture, Sports,
Science, and Technology (MEXT)-Supported Program for the Strategic
Research Foundation at Private Universities "Topological Science" (Grant No. S1511006)
and in part by the Japan Society for the Promotion of Science (JSPS)
Grant-in-Aid for Scientific Research (KAKENHI) Grant Number 
18H01217,
18H05406, and 18H01211. 

\end{document}